\documentclass[12pt]{article}
\usepackage{amsmath}
\usepackage{amscd}
\usepackage{amsfonts}
\usepackage{amsthm}
\usepackage{amssymb}
\usepackage{bm}
\newcommand{\mathbbone}{{\rm 1\mskip-4.5mu l}}
\newcommand{\dd}{d}

\begin{document} 

\bibliographystyle{prsty}

\title{{\bf The Orbifolds of Permutation-Type as \\
Physical String Systems \\
at Multiples of $\mathbf{c=26}$\\
IV. Orientation Orbifolds Include Orientifolds}}
\author{M.~B. Halpern\footnote{halpern@physics.berkeley.edu}\\
Department of Physics, 
University of California, \\
and Theoretical Physics Group, \\
Lawrence Berkeley National Laboratory\\
University of California, 
Berkeley, California 94720, USA}

\maketitle

\begin{abstract}

In this fourth paper of the series, I clarify the somewhat mysterious 
relation between the large class of {\it orientation orbifolds} (with twisted open-string
CFT's at $\hat c=52$) and {\it orientifolds} (with untwisted open strings
at $c=26$), both of which have been associated to division by world-sheet 
orientation-reversing automorphisms. In particular -- following a spectral clue in the previous 
paper -- I show that, even as an {\it interacting string
system}, a certain half-integer-moded orientation orbifold-string 
system is in fact
equivalent to the archetypal orientifold. The subtitle of this paper, 
 that orientation orbifolds include and generalize standard orientifolds,
then follows because there are many other orientation orbifold-string 
systems -- with
higher fractional modeing -- which are not equivalent to 
untwisted string systems.
\end{abstract}

\clearpage
\tableofcontents

\renewcommand{\thesection}{\arabic{section}.}
\renewcommand{\theequation}{\thesection\arabic{equation}}
\setcounter{equation}{0}
\setcounter{section}{0}

\newpage 
\section{Introduction\label{sec1}}

Non-abelian and free-bosonic {\it orientation orbifolds} [12,13,15] were first 
studied as a large class of examples in the orbifold program 
[1-11,12-15], which attempts to
construct all orbifold CFT's. From the beginning there has been some mystery
about the relation of the orientation orbifolds to ordinary 
{\it orientifolds} [16], since both classes ostensibly arise in the division 
$A(H_{-})/H_{-}$ of a CFT $A(H_{-})$ by a symmetry group $H_{-}$ which 
contains world-sheet orientation-reversing automorphisms. However, 
following standard methods in orbifold theory, the
orientation-orbifold CFT's contain twisted open-string sectors 
(corresponding to the orientation-reversing automorphisms) at 
{\it twice} the central charge $c$ of $A(H_{-})$ -- as well as an equal number 
of twisted closed-string sectors (which form an ordinary space-time 
orbifold) at $c$. By twisted I mean fractionally-moded, as expected 
in the twisted sectors of any orbifold CFT. A simple understanding of 
the doubled central charge of the twisted open-string 
 sectors is found in Ref. [12].

In the present series of papers [17-19], we have been considering 
the critical orbifolds of permutation-type as candidates for new {\it physical string systems} at higher 
central charge, including the space-time permutation orbifolds at 
any multiple of $c=26$, as well as
the general free-bosonic orientation orbifold:
\begin{equation}
\label{eq1.1}
\frac{U(1)^{26}}{H_-}\,,\quad H_-=\mathbb{Z}_2 (\text{world sheet})\times H
\end{equation}
In these cases, the non-trivial element of $\mathbb{Z}_{2}(\mbox{w.s.})$
permutes the left- and right-movers of the critical closed string $U(1)^{26}$
while the extra automorphisms $H$ act uniformly on both chiralities. 
The twisted open-string sectors of these orbifold CFT's  
show an essentially-unbounded variety of fractional modeing at central charge $\hat c=52$.

As {\it string theories} however, the orbifolds of permutation-type 
must also satisfy certain {\it extended physical-state conditions} 
[16-18], which 
restrict their physical spectra relative to the underlying orbifold 
CFT's, and this has led to some surprises. In particular, although the 
$\hat c=52$ orbifold-string spectra [18] are generically unfamiliar, we have learned 
that the spectra of some of the simplest (half-integer-moded) closed or open $\hat c=52$ strings 
are equivalent to those of ordinary untwisted closed or 
open strings at $c=26$.

In the present paper, I will follow one of these spectral clues  to  
clarify the relation between  orientation orbifolds and 
orientifolds.

\vspace{.05in}\noindent 1) In the first place, I will study here only the simple 
two-sector, half-integer-moded orientation orbifold
\begin{equation}
\label{eq1.2}
\frac{U(1)^{26}}{H_-}\,,\quad H_-=\left(1;\, \tau_-\times(-1)\right)
\end{equation}
in further detail, showing that -- even as an {\it interacting string 
theory} -- this orientation orbifold is equivalent to the archetypal orientifold  
\begin{subequations}
\begin{align}
\label{eq1.3a}
\sigma&=0:\quad\text{unoriented closed string at } c=26 \\
\label{eq1.3b}
\sigma&=1:\quad\text{twisted open string at } \hat c=52 \\
&\quad\quad\qquad\,=\text{ordinary $NN$ string at } c=26
\nonumber
\end{align} 
\end{subequations}
that is, the conventional open-closed bosonic string system [20].
In the course of this discussion, we will see the extended 
physical-state conditions of Refs. [17-19] realized via {\it extended 
Ward identities} in the twisted-tree diagrams of the orbifold-string 
system -- and our unconventional, orientation-orbifold description of the conventional 
system also gives a new and very 
simple derivation of the ratio of open- to closed-string Regge 
slopes.

\vspace{.05in}\noindent 2) A second conclusion borders on the philosophical: In 
contrast with the language of orientifolds , our construction clearly
shows the relation of the conventional open-closed 
string system to standard orbifold conformal field theory [21-25,1-15].  In particular, the twisted 
sectors of any orbifold CFT must show fractional modeing, and indeed
-- although it is not visible in mass-shell emission from the  boundaries -- half-integer 
modeing is still present in the  bulk of our twisted open string 
(1.3b).

\vspace{.05in}\noindent 3) The final conclusion is that, as string theories, orientation orbifolds 
include orientifolds
\begin{equation}
\label{eq1.4}
\{\text{orientation orbifolds}\}\, \supset\, \{\text{orientifolds}\}.
\end{equation}
This embedding follows from the equivalence discussed here for the half-integer-moded case (1.2), and the fact that
there exist many other 
orientation orbifolds (shown in Eq. (1.1)) with higher fractional modeing
whose critical orbifold-string systems are {\it not equivalent} [19] to untwisted strings.

\vspace{.05in}It is hoped that the computations given here will serve as a 
prototype for studying these more general orbifold-string systems at 
the interacting level. I remind that all the $\hat c=52$ 
orbifold-string spectra have an equivalent, unconventionally-twisted 
$c=26$ description [19] of the twisted $\hat c=52$ matter. Indeed, our 
result here extends an example of this spectral equivalence to the interacting 
level, and the general $c=26$ spectral equivalence strongly 
suggests that an equivalent but generically-unconventional $c=26$ description 
may also exist for all these interacting theories.

\setcounter{equation}{0}
\section{A Simple Orientation Orbifold\label{sec2}}
To establish notation, I begin with a few well-known facts about the 
ordinary (decompactified) critical closed bosonic string $U(1)^{26}$:
\begin{subequations}
\begin{gather}
\label{eq2.1a}
L(0)=-\frac{J^2(0)}{2}+R,\qquad \bar L(0)=-\frac{\bar J^2(0)}{2}+\bar R\\
\label{eq2.1b}
\bar R=R, \qquad J_\mu(0)=\bar J_\mu(0) \simeq\, T_\mu,\qquad
\mu=0,1,\ldots,25\\
\label{eq2.1c}
A\cdot B=A_\mu \eta^{\mu\nu} B_\nu,\qquad
\eta=
\begin{pmatrix} 
1 & 0 \\
0 & -\mathbbone 
\end{pmatrix}.
\end{gather}
\end{subequations}
Here $\eta^{\boldsymbol{\cdot}}$ is the inverse target-space metric, 
the zero-mode eigenvalues $\{T\}$ 
are the dimensionless momenta and the relations in Eq. (2.1b) are 
consequences of level-matching $L(0)=\bar{L}(0)$. The relation to the dimensionful 
momenta $\{k\}$ is 
\begin{equation}
\label{eq2.2}
T_\mu=\sqrt{\alpha'_c}\,k_\mu
\end{equation}
where $\alpha'_c$ is the ordinary closed-string Regge slope. To eliminate 
negative-norm states, one must fix $L(0)=\bar{L}(0)=1$ and hence
$T^{2}=-2$ for the closed-string ground state, but I will relax this 
mass-shell condition during our discussion of the orientation orbifold
as a conformal-field-theoretic system.

The particular orientation orbifold we will study in detail is 
\begin{equation}
\label{eq2.3}
\frac{U(1)^{26}}{H_-}\,,\quad H_-=\left(1;\,\tau_-\times(-1)\right)
\end{equation}
where $\tau_{-}$ is world-sheet orientation reversal. The two sectors 
$\sigma$ of this orbifold correspond to the following automorphic action on the 
untwisted current modes of $U(1)^{26}$
\begin{subequations}
\begin{align}
\label{eq2.4a}
\sigma&=0:\quad J_\mu(m)'=J_\mu(m),\quad \bar J_\mu(m)'=\bar J_\mu(m)\\
\label{eq2.4b}
\sigma&=1:\quad J_\mu(m)'=-\bar J_\mu(m),\quad \left.\bar 
J_\mu(m)\right. \!'=- J_\mu(m)
\end{align} 
\end{subequations}
and the corresponding orientation-orbifold sectors, constructed by standard 
orbifold methods  from $U(1)^{26}$, are similarly labeled:
\begin{equation}
\label{eq2.5}
\begin{split}
\sigma&=0:\quad\text{untwisted (symmetric) sector at } c=26 \\
\sigma&=1:\quad\text{twisted sector at } \hat c=52. 
\end{split}
\end{equation}
In particular, the untwisted sector is an unoriented closed string, 
while the twisted sector is  the half-integer-moded open-string
conformal field theory discussed in the following section. 

\setcounter{equation}{0}
\section{The Twisted Sector as an Open-String CFT\label{sec3}}
 The twisted open-string CFT of the
 orientation orbifold (2.3) was constructed from the closed-string 
 sector $\{\mbox{untwisted closed}\to\mbox{twisted open}\}$ among 
 the large class of free-bosonic examples in Refs. 
[12,13,15], and further discussed in Refs. [17,19]. 
More specifically, we may use the metric relation 
 $G_{\mu\nu}=-{\eta}_{\mu\nu}$ 
to read off many results from Ref. [15], 
beginning with the action, the mode expansions and the orbifold
Virasoro generators of this particular twisted sector:
\begin{subequations}
\begin{gather}
\label{eq3.1a}
\hat S=-\frac{1}{4\pi} \eta^{\mu\nu}\int \dd t\int_0^\pi \dd\xi
\sum_{u=0}^1\partial_+ \hat x^{1\mu u}\, \partial_- \hat x^{1\nu,-u}\\
\label{eq3.1b}
\partial_{\pm}=\partial_t\pm\partial_\xi\\
\label{eq3.1c}
\text{DN}:\quad \hat x_{1\mu 0}(\xi,t)=2
\sum_{m\in {\mathbb Z}}\frac{\hat J_{1\mu 0}(m+\tfrac{1}{2})}{m+\tfrac{1}{2}}
e^{-i(m+1/2)t}\sin\left((m+\tfrac{1}{2})\xi\right)\\
\label{eq3.1d}
\text{NN}:\quad \hat x_{1\mu 1}(\xi,t)=\hat q_{1\mu 1}+
2\hat J_{1\mu 1}(0)t+2i\sum_{m\neq 0}\frac{\hat J_{1\mu 1}(m)}{m}
e^{-imt}\cos(m\xi)
\end{gather}
\begin{multline}
\label{eq3.1e}
\hat L_u\left(m+\tfrac{u}{2}\right)=\tfrac{13}{8}\delta_{m+\tfrac{u}{2},0}\\
-\tfrac{1}{4}\eta^{\mu\nu}\sum_{v=0}^1\sum_{p\in{\mathbb Z}}
:\hat J_{1\mu v}\left(p+\tfrac{v+1}{2}\right)\hat J_{1\nu, u-v}
\left(m-p+\tfrac{u-v-1}{2}\right):_M.
\end{multline}
\end{subequations}
In these results, $:\cdot:_{M}$ is standard mode normal-ordering, all 
quantities are periodic $u\to u\pm2$ and $\bar u=0,1$ is the 
pullback to the fundamental region. Up and down indices are simply 
related
\begin{equation}
\label{eq3.2}
\hat A^{\mu u}=\tfrac{1}{2}\eta^{\mu\nu}\hat A_{\nu,-u}=
\tfrac{1}{2}\eta^{\mu\nu}\hat A_{\nu u}
\end{equation}
where the last form holds by periodicity in $u$. Note 
that this open-string sector contains $26$ integer-moded $NN$ degrees of freedom and 
$26$ half-integer-moded \footnote{Half-integer-moded scalar fields 
[21] (and the corresponding twisted open strings with ND or DN 
boundary conditions [22,23]) provided the first examples of twisted sectors 
of orbifolds.} $DN$ degrees of freedom, giving a total 
central charge $\hat c=52$. Half-integer moding is of course the 
standard twisting of an order-two automorphism such as that in Eq. 
(2.4b), and a simple understanding of the doubled central charge of 
all open-string orientation-orbifold sectors is given in 
Ref. [12].

The twisted algebras \footnote{The full quasi-canonical algebra, 
branes and twisted non-commutative geometry of 
the general free-bosonic orientation orbifold is given in 
Refs. [12,15].} of this sector are
\begin{subequations}
\begin{gather}
\label{eq3.3a}
\left[ \hat J_{1\mu u}\left(m+\tfrac{u+1}{2}\right),
\hat J_{1\nu v}\left(n+\tfrac{v+1}{2}\right)\right]
=-2\left(m+\tfrac{u+1}{2}\right) \eta_{\mu \nu}\delta_{m+n+\tfrac{u+v}{2}+1,0}\\
\label{eq3.3b}
\left[ \hat L_u(m+\tfrac{u}{2}),\hat J_{1\mu v}(n+\tfrac{v+1}{2})\right]
=-\left(n+\tfrac{v+1}{2}\right) 
\hat J_{1\mu,u+v}\left(m+n+\tfrac{u+v+1}{2}\right)\\
\label{eq3.3c}
\begin{split}
\left[\hat L_u\left(m+\tfrac{u}{2}\right), 
\hat L_v\left(n+\tfrac{v}{2}\right)\right]&=
(m-n+\tfrac{u-v}{2})\hat L_{u+v}\left(m+n+\tfrac{u+v}{2}\right)\\
&+\tfrac{52}{12}\left(m+\tfrac{u}{2}\right)\left( (m+\tfrac{u}{2})^2-1\right)
\delta_{m+n+\tfrac{u+v}{2},0}
\end{split}
\end{gather}
\end{subequations}
where Eq. (3.3c), which appears universally in all $\hat c=52$ 
orbifold strings,  is called an order-two orbifold Virasoro algebra [1,26,9]. 
The generators $\{\hat{L}_{0}(m)\}$ of the integral 
Virasoro subalgebra, with central charge $\hat c=52$, are the 
physical Virasoro generators of the sector. Note also the presence of 
two sets of time-like modes $\{\hat{J}_{10u}, \bar u =0,1\}$ and 
hence twice the conventional number of negative-norm states in the CFT.
As string theories however, we know that all twisted open and closed $\hat c=52$
strings must also satisfy the so-called {\it extended physical-state conditions} [18,19]
\begin{equation}
\label{eq3.4}
\left(\hat L_u\left( (m+\tfrac{u}{2})\geq 0\right) -
\tfrac{17}{8}\delta_{m+\tfrac{u}{2},0}\right)\vert \chi\rangle=0,
\quad \bar u=0,1
\end{equation}
leaving a physical spectrum which in this particular case [19] is identical to that of 
an ordinary untwisted open $NN$ string at $c=26$. 
The extended physical-state conditions (3.4) are the 
operator-analogues of the extended Polyakov constraints of 
$\mathbb{Z}_{2}$-twisted permutation gravity [17] and, more precisely, 
these conditions are a  consequence of the 
twisted BRST systems [18] of $\hat c=52$ matter. We will see 
another derivation of these conditions for $(m+\tfrac{u}{2})>0$ at 
the interacting level below.

To study the interacting theory we will also need the twisted 
Minkowski-space vertex operator
\begin{equation}
\label{eq3.5}
\hat{\hat{g}}_+(T,\xi,t)\propto\,\,\, : 
e^{-\frac{i}{2}\left(\hat x_{1\mu 0}(\xi,t)+\hat x_{1\mu 1}(\xi,t)\right)
\eta^{\mu\nu}T_\nu}:
\end{equation}
which is an example of the free-bosonic or abelian ``limit'' of the twisted 
affine primary fields [6-15] of the non-abelian orientation orbifolds.
According to the orbifold construction $\{\mbox{closed string} \to \mbox{twisted open 
string}\}$ in Refs. [12,13,15], the quantities $\{T\}$ in Eq. (3.5) are the {\it 
same} untwisted closed-string momenta $\{T\}$ defined for the 
ordinary closed string $U(1)^{26}$
in Eqs. (2.1) and (2.2). The vertex operator given here is in fact 
only one of two irreducible components of the vertex operator 
derived in Refs. [12,15], the latter including an extra $2\times2$ reducible matrix 
structure which is further discussed in Appendix A.

I limit the discussion here to the twisted vertex operator for emission at 
$\xi=\pi$ (returning later to emission at $\xi=0$). Defining the 
quantities
\begin{equation}
\label{eq3.6}
z\equiv e^{it},\quad A\cdot B = A_\mu\eta^{\mu\nu} B_\nu
\end{equation}
we find that the precise form \footnote{See in particular Subsec. 5.4 
of Ref. [15].} of this emission vertex is
\begin{subequations}
\begin{equation}
\label{eq3.7a}
\hspace*{-2.9in}\hat g_+(T,\pi,z)\equiv\, z^{T^2}\hat{\hat g}_+(T,\pi,z)
\end{equation}
\begin{multline}
\label{eq3.7b}
 \hspace*{.1in}= \,z^{T^2}e^{-i\sqrt{2}T\cdot q}e^{-\sqrt{2}\ln z\, T\cdot 
 J(0)}\,\times\\
\times \exp\left[-\sqrt{2}T\cdot\sum_{m\ge 1}\frac{J(-m)}{m}(-1)^m z^m\right]
\exp\left[\sqrt{2}T\cdot\sum_{m\ge 1}\frac{J(m)}{m}(-1)^m z^{-m}\right]\\
\times \exp\left[-i T\cdot \sum_{m\le -1}
\frac{\hat J_{10}(m+\tfrac{1}{2})}{m+\tfrac{1}{2}}(-1)^m z^{-(m+\tfrac{1}{2})}
\right]\\
\times
\exp\left[-iT\cdot \sum_{m>0}\frac{\hat J_{10}(m+\tfrac{1}{2})}{m+\tfrac{1}{2}}
(-1)^m z^{-(m+\tfrac{1}{2})}\right].
\end{multline}
\end{subequations}
For computational convenience, I have here reexpressed the 
integer-moded $NN$ subsystem in terms of conventional 
Fubini-Veneziano operators [27]
\begin{subequations}
\begin{gather}
\label{eq3.8a}
J_\mu(m)\equiv \frac{1}{\sqrt{2}}\hat J_{1\mu 1}(m),\quad
q_\mu\equiv \frac{1}{2\sqrt{2}}\hat q_{1\mu 1}\\
\label{eq3.8b}
\left[q_\mu, J_\nu (0)\right]=-i\eta_{\mu\nu},\quad
\left[J_\mu(m), J_\nu (n)\right]=-m\eta_{\mu\nu}\delta_{m+n,0}
\end{gather}
\end{subequations}
but I keep the original notation for the commutators with the vertex 
operator:
\begin{subequations}
\begin{gather}
\label{eq3.9a}
\left[\hat J_{1\mu u}\left(m+\tfrac{u+1}{2}\right), \hat g_+(T,\pi,z)\right]=
e^{i\pi(m+\tfrac{u+1}{2})} 2 T_\mu z^{m+\tfrac{u+1}{2}}\hat g_+(T,\pi,z)\\
\label{eq3.9b}
\left[\hat L_u\left(m+\tfrac{u}{2}\right),\hat g_+(T,\pi,z)\right]=
e^{i\pi(m+\tfrac{u}{2})}\left( z \partial_z + (m+1+\tfrac{u}{2})4\Delta(T)\right)
\hat g_+(T,\pi,z)\\
\label{eq3.9c}
\Delta(T)\equiv -\frac{T^2}{2}.
\end{gather}
\end{subequations}
The phases in Eqs. (3.7) and (3.9) are a consequence of the choice 
$\xi=\pi$ in the twisted string cooordinates (3.1c,d).

Finally, we will need the following properties of the momentum-boosted 
twist-field states
\begin{subequations}
\begin{gather}
\label{eq3.10a}
\vert T\rangle \equiv \lim_{z\to 0} z^{-T^2} \hat g_+(T,\pi,z)\vert 0\rangle\\
\label{eq3.10b}
\left( J_\mu (m\ge 0)-\sqrt{2} T_\mu \delta_{m,0}\right)\vert T\rangle
= J_{1\mu 0}\left( (m+\tfrac{1}{2}) > 0\right) \vert T\rangle =0\\
\label{eq3.11b}
\left( \hat L_u\left((m+\tfrac{u}{2})\ge 0\right)-
\left(\tfrac{13}{8}+2\Delta(T)\right) \delta_{m+\tfrac{u}{2},0}\right)\vert 
T\rangle=0
\end{gather}
\end{subequations}
one of which will be selected below as the ground state of the 
twisted open string.  

\setcounter{equation}{0}
\section{Twisted Tree Graphs\label{sec4}}

I turn now to discuss the twisted $\hat c=52$ open-string CFT of the 
previous section as a sector of the full {\it interacting} 
orientation-orbifold string theory.

Using the quantities introduced in the previous section, I define the
 n-point {\it twisted tree graphs} of the interacting string theory by the 
following ``sidewise construction'':
\begin{subequations}
\begin{gather}
\label{eq4.1a}
\begin{split}
\hat A_n(\{ T\})&\equiv \langle -T^{(n)}\vert\hat g_+(T^{(n-1)},\pi,1)
\hat D[\hat L_0(0)]\hat g_+(T^{(n-2)},\pi,1) \cdots \\
& \quad \quad \quad\quad\quad\hspace{1in} \cdots \hat D[\hat L_0(0)] \hat g_+(T^{(2)},\pi,1)\vert T^{(1)}\rangle
\end{split}\\
\label{eq4.1b}
\hat D[\hat L_0(0)]=\frac{1}{2(\hat L_0(0)-\hat a_2)}
=\frac{1}{2}\int_0^1 \dd x\, x^{\hat L_0(0)-\hat a_2-1}\\
\label{eq4.1c}
\hat a_2=\frac{17}{8}.
\end{gather}
\end{subequations}
Here $\hat{g}_{+}$ is the twisted vertex operator (3.7) for emission at 
$\xi=\pi$, now additionally evaluated at $z=1$. In the twisted 
propagator $\hat D$, the operator $\hat{L}_{0}(0)$ is the zero mode 
of the integral Virasoro subalgebra, and the constant $\hat{a}_{2}$ was determined 
from the $m=u=0$ component of the extended physical-state condition 
in Eq. (3.4). In the sidewise construction, one sees the $\hat c=52$ twisted open-string 
sector running horizontally (sidewise) in the direct channel, whereas 
the behavior of the twisted trees in the cross channels is not yet visible 
\footnote{The sidewise construction may be unfamiliar today, though it was well-known in the first 
string era. Indeed it was this construction which was used to obtain 
the (sidewise) $R\to NS$ amplitudes [28] and, with the addition 
of twisted scalar fields, the (sidewise) twisted sector $\to$ untwisted 
sector amplitudes [21] in early orbifold theory.}.

Comparing the propagator $\hat D$ with the properties of the 
twist-field states in Eq. (3.10), one then finds 
that the ground state and hence the twisted vertex operators must 
satisfy the mass-shell conditions
\begin{equation}
\label{eq4.2}
\Delta(T^{(i)})=\frac{1}{4},\quad
(T^{(i)})^2=-\frac{1}{2},\quad
i=1,\ldots,n
\end{equation}
so that the ground state is a pole of the propagator. Moreover, these 
conditions and the commutator (3.9b) with the extended Virasoro 
generators imply the following 
further properties of the twisted vertex operators 
\begin{subequations}
\begin{gather}
\label{eq4.3a}
\hat g_+(T,\pi,z)=z^{\hat L_0(0)}\hat g_+(T,\pi,1)z^{-\hat L_0(0)-1}\\
\label{eq4.3b}
\left[\Big(e^{-i\pi(m+\tfrac{u}{2})}\hat L_u(m+\tfrac{u}{2})-\hat 
L_0(0)\Big),
\hat g_+(T,\pi,1)\right]=(m+\tfrac{u}{2})\hat g_+(T,\pi,1)
\end{gather}
\end{subequations}
where Eq. (4.3b) is a generalization of the so-called stability 
condition [29] in untwisted string theory.

\setcounter{equation}{0}

\section{Extended Ward Identities at $\mathbf{\hat c=52}$\label{sec5}}

It was conjectured in Ref. [18] that the extended physical-state 
conditions in Eq. (3.4) would also follow from extended Ward identities in the
interacting theory.

To see this explicitly in the present example, I begin by 
defining the extended (twisted) gauge operators: 
\begin{subequations}
\begin{gather}
\label{eq5.1a}
\hat W_u(m+\tfrac{u}{2})\,\equiv \,e^{-i\pi(m+\tfrac{u}{2})}
\hat L_u(m+\tfrac{u}{2})-\left( \hat L_0(0)+m+\tfrac{u}{2}-\hat 
a_2\right)\\
\label{eq5.1b}
 \bar u=0,1.
\end{gather}
\end{subequations}
In order to see that these gauges are active in the twisted trees 
(4.1), the 
following vertex and propagator identities are helpful
\begin{subequations}
\begin{gather}
\label{eq5.2a}
\hat W_u(m+\tfrac{u}{2})\hat g_+(T^{(i)},\pi,1) =\hat g_+(T^{(i)},\pi,1)
\left(e^{-i\pi(m+\tfrac{u}{2})}\hat L_u(m+\tfrac{u}{2}) -\hat L_0(0)+\hat a_2\right)\\
\label{eq5.2b}
\hspace*{-1.2in}\left(e^{-i\pi(m+\tfrac{u}{2})}\hat L_u(m+\tfrac{u}{2})-
\hat L_0(0)+\hat a_2\right)\hat D[\hat L_0(0)]=\\
\hspace{2in}=\hat D\!\left[\hat L_0(0)+m+\tfrac{u}{2}\right]\hat 
W_u(m+\tfrac{u}{2}) {\nonumber}
\end{gather}
\end{subequations}
where Eqs. (5.2a) and (5.2b) follow respectively from the extended 
stability condition (4.3b) and the orbifold Virasoro algebra (3.3c).

Then we find after some algebra the {\it extended Ward identities} at $\hat c=52$:
\begin{subequations}
\begin{gather}
\label{eq5.3a}
\hat W_u\left( (m+\tfrac{u}{2}) > 0\right)\hat g_+(T^{(m)},\pi,1)
\hat D[\hat L_0(0)]\cdots\hat D[\hat L_0(0)]\hat g_+(T^{(1)},\pi,1)
\vert \chi\rangle =0\\
\label{eq5.3b}
\forall\, \vert \chi\rangle\,\,\text{s.t.}\,\,
\left(\hat L_u\left((m+\tfrac{u}{2})\ge 0\right)-
\hat a_2\delta_{m+\tfrac{u}{2},0}\right)\vert \chi\rangle=0,
\quad \bar u=0,1.
\end{gather}
\end{subequations}
It should be emphasized that  Eq. (5.3b) is the {\it same} extended (twisted) physical-state condition (3.4)
obtained from the general twisted $\hat c=52$ BRST system in Ref. [18]. In particular,
Eqs. (3.10c) and (4.2) tell us that the twisted open-string ground state at 
$T^{2}=-1/2$ is a physical state:
\begin{subequations}
\begin{gather}
\label{eq5.4a}
\vert T\rangle=\lim_{z\to 0} z^{\frac{1}{2}}\hat g_+(T,\pi,z)\vert 0\rangle,
\quad
\langle -T \vert=\lim_{z\to\infty}
\langle 0\vert\, z^{\frac{3}{2}}\hat g_+(T,\pi,z)\\
\label{eq5.4b}
\left(\hat L_u\left((m+\tfrac{u}{2})\ge 0\right)-
\hat a_2\delta_{m+\frac{u}{2},0}\right)\vert T\rangle=
\langle -T\vert \left( \hat L_u\left((m+\tfrac{u}{2})\le 0\right)
-\hat a_2 \delta_{m+\frac{u}{2},0}\right){\nonumber}\\
=0. 
\end{gather}
\end{subequations}
Unconventional prefactors in asymptotic conditions, such as those
in Eqs. (3.10) and (5.4a), are well-known in the orbifold program 
(see e.g. Ref. [9]).

I also remind that the solution of the extended physical-state 
condition (5.3b) with the extended Virasoro generators (3.1e)
is known [19], showing that the physical spectrum of this {\it particular} twisted open 
$\hat c=52$ string is the {\it same} as that of an ordinary untwisted open
NN string at $c=26$. This leads us to suspect with Ref. [19] that 
the twisted trees (4.1) of the orientation orbifold may be 
an unconventional realization of the tree graphs of ordinary NN strings.

\setcounter{equation}{0}
\section{Evaluation of the Twisted Trees\label{sec6}}

To evaluate the twisted trees, I begin with the n-point correlators
of the open-string orientation-orbifold CFT
\begin{subequations}
\begin{multline}
\label{eq6.1a}
\langle 0\vert \hat g_+(T^{(n)},\pi,z_n)\cdots\cdot\hat g_+(T^{(1)},\pi,z_1)
\vert 0\rangle\\
\,\,\,\,=\delta^{26}\left(\sum_{i=1}^n T^{(i)}\right)\prod_{j=1}^n z_i^{-\frac{1}{2}}
\sum_{i<j}\left\{z_i(1-\frac{z_j}{z_i})\left(
\frac{1-(z_j/z_i)^{1/2}}{1+(z_j/z_i)^{1/2}}\right)\right\}^{-2T^{(i)}\cdot T^{(j)}}
\end{multline} 
\begin{equation}
\label{eq6.1b}
=\delta^{26}\left(\sum_{i=1}^n T^{(i)}\right)
\prod_{j=1}^n z_i^{-\tfrac{1}{2}}\sum_{i<j}
\left(\sqrt{z_i}-\sqrt{z_j}\right)^{-4 T^{(i)}\cdot T^{(j)}}
\end{equation}
\end{subequations}
which follow from the twisted vertex operators (3.7) when 
$\forall\, \Delta(T^{(i)})=1/4$ (see Eq. (4.2). Then we may use the integral 
representation (4.1b) of the twisted propagator to evaluate
the sidewise construction in Eq. (4.1) as follows:
\begin{subequations}
\begin{multline}
\label{eq6.2a}
\hat A_n(\{T\})=2^{-(n-3)}\int_0^1
\prod_{i=2}^{n-2}(\dd x_i\, x_i^{-\hat a_2-1})\,\times\\
 \times\langle -T^{(n)}\vert \hat g_+(T^{(n-1)},\pi,1)
x_{n-2}^{\hat L_0(0)}\cdots x_1^{\hat L_0(0)}\hat g_+(T^{(2)},\pi,1)
\vert T^{(1)}\rangle
\end{multline}
\begin{multline}
\label{eq6.2b}
=2^{-(n-3)}\int_0^1 \dd z_{n-2}\,\int_0^{z_{n-2}}\dd z_{n-3}\,
\cdots\int_0^{z_3}\dd z_2 \,\times\\
\times \lim_{\substack{z_n\to\infty\\
z_{n-1}\to 1\\
z_1\to 0}}
z_1^{1/2}z_n^{3/2}\langle 0\vert \hat g_+(T^{(n)},\pi,z_n)\cdots
\hat g_+(T^{(1)},\pi,z_1)\vert 0\rangle
\end{multline}
\begin{multline}
\label{eq6.2c}
=2^{-(n-3)}\delta^{26}\left(\sum_{i=1}^n T^{(i)}\right)
\int_0^1 \dd z_{n-2}\,\int_0^{z_{n-2}}\dd z_{n-3}\cdots\int_0^{z_3}\dd 
z_2\,\times\\
\times \prod_{i=2}^{n-2} z_i^{-2T^{(i)}\cdot T^{(1)}-\tfrac{1}{2}}
(1-\sqrt{z_i})^{-4 T^{(n-1)}\cdot T^{(i)}}
\prod_{i<j}\left(\sqrt{z_i}-\sqrt{z_j}\right)^{-4 T^{(i)}\cdot T^{(j)}}.
\end{multline}
\end{subequations}
Here I have used the boost (4.3a) and the asymptotic 
relations (5.4a), as well as the standard change of variables
$\{ z_i\equiv \prod_{j=i}^{n-2}x_j\}$ to obtain Eq. (6.2b) -- and the  
correlators (6.1) to obtain the last form. The roots in all these 
expressions reflect the half-integer modeing of the open-string 
orientation-orbifold CFT.

But now consider the non-linear change of variables
\begin{equation}
\label{eq6.3}
z_i\equiv u_i^2,\quad i=2,\ldots,n-2
\end{equation}
which puts our result in the final form:
\begin{multline}
\label{eq6.4}
\hat A_n(\{T\})=\delta^{26}\left(\sum_{i=1}^n T^{(i)}\right)
\int_0^1 \dd u_{n-2}\,\int_0^{u_{n-2}}\dd u_{n-3}\cdots
\int_0^{u_3}\dd u_2\,\times\\
\times \prod_{i=2}^{n-2}
\left( u^{-4 T^{(i)}\cdot T^{(1)}} (1-u_i)^{-4 T^{(n-1)}\cdot T^{(i)}}\right)
\prod_{i<j}(u_i-u_j)^{-4 T^{(i)}\cdot T^{(j)}}.
\end{multline}
The non-linear transformation has removed all the roots, and indeed, under 
the following 
identification of the ordinary, untwisted open-string Regge slope ${\alpha}_{0}'$ and the 
dimensionful momenta $\{k\}$
\begin{equation}
\label{eq6.5}
\sqrt{2}T^{(i)}=\sqrt{\alpha'_0}k_i,\quad
\alpha'_0 k_i^2=-1,\quad i=1,\ldots,n
\end{equation}
we see that the twisted $\hat c=52$ trees in Eq. (4.1) are nothing but a 
new factorization of the trees of the ordinary untwisted open NN 
string [30] at $c=26$\,!

The same NN trees are found as well for multiple emission from the twisted open 
string at  $\xi=0$, 
where the half-integer-moded DN coordinates in Eq. (3.1c) are 
trivially suppressed.  These evaluations also exhibit the 
expected [17-19] no-ghost theorem (including the decoupling of zero-norm states) for
the twisted $\hat c=52$ sector of 
this orientation orbifold.

\setcounter{equation}{0}
\section{Open- and Closed-String Regge Slopes\label{sec7}}

As a byproduct of our computation
\begin{equation}
\label{eq7.1}
 \mbox{closed}  \,(c=26) \,\to \,\mbox{twisted 
open}\, (\hat c=52) \,\simeq \,\,
\mbox{untwisted open} \,(c=26)
\end{equation}
 we also obtain a simple new 
derivation of the ratio of the Regge slope $\alpha_{c}'$ of the 
original untwisted closed string to the slope $\alpha_{0}'$ of the resulting 
untwisted open string. The key is the appearance of the {\it same} 
[12,13,15] {\it dimensionless 
momenta} $\{T\}$ in both the untwisted and the twisted sector of the orientation 
orbifold \footnote{This phenomenon is quite general in the orbifold 
construction of twisted sectors from the untwisted sector, so that e.g. the untwisted
representation matrices $\{T\}$ of Lie g appear in the twisted 
representation matrices $\{\mathcal{T}(T)\}$ of the twisted sectors of WZW 
orbifolds [6-11] and WZW orientation orbifolds 
[12,13,15]}.

In fact, we need only combine Eqs. (2.2) and (6.5) to find the correct 
relation [31] between the slopes 
\begin{subequations}
\begin{gather}
\label{eq7.2a}
\sqrt{\alpha'_c}\,k=T=\sqrt{\frac{\alpha'_0}{2}}\, k\\
\label{eq7.2b}
\longrightarrow\quad \alpha'_0=2\alpha'_c
\end{gather}
\end{subequations}
where $\alpha(s)=\alpha(0)+\alpha' s$ is the form of either leading 
trajectory. Here I have assumed only that the dimensionful momenta
$\{k\}$ are the same for both string types -- which is required by 
momentum conservation in any open-closed string interaction. I also 
emphasize that our computation involves going off-shell
\begin{equation}
\label{eq7.3}
\text{closed:}\quad T^2=-2\quad\longrightarrow\quad
\text{open:}\quad T^2=-\frac{1}{2}
\end{equation}
in order to pass between the ground states of the two strings. This 
reflects the fact that the orientation-orbifold construction $\{ 
\mbox{untwisted closed 
string} \to 
 \mbox{twisted open string} \}$ is fundamentally conformal-field-theoretic.

\setcounter{equation}{0}
\section{Conclusions\label{sec8}}

We have confirmed at the interacting level the conclusion reached for 
the physical spectrum in Ref. [19]: As a string system, the simple 
orientation orbifold
\begin{equation}
\label{eq8.1}
\frac{U(1)^{26}}{H_-}\,,\quad
H_-=(1,\,\tau_1\times(-1))
\end{equation}
is equivalent to the archetypal orientifold
\begin{subequations}
\begin{align}
\label{eq8.2a}
\sigma&=0:\quad\text{unoriented closed string at } c=26 \\
\label{eq8.2b}
\sigma&=1:\quad\text{twisted open string at } \hat c=52 \\
&\quad\quad\qquad=\text{ordinary $NN$ string at } c=26.
\nonumber
\end{align} 
\end{subequations}
that is, the ordinary open-closed bosonic string system.
Indeed, the evaluation of the twisted trees in Sec. 6 has 
sharpened the equivalence even at the spectral level -- showing now 
the correct decoupling of null physical states.
 As a bonus, our 
unconventional formulation of the conventional system also provided
a new derivation of the ratio (7.2b) of open- to 
closed-string Regge slopes.

This clarifies the somewhat mysterious relation between 
the orientation orbifold (8.1) and the orientifold --
both of whose open-string CFT's have been differently associated to division by 
the world-sheet orientation-reversing automorphism $(-\tau_{-})$. The resolution is that, {\it as mass-shell string 
 theories}, 
these particular two open-string sectors are identical, even at the interacting 
level.

Although we have found equivalence in this case for the spectrum and mass-shell emissions at 
the string boundaries, I emphasize that the orientation-orbifold picture
remains {\it qualitatively different} than the conventional picture in other 
regions of the theory, including  
\begin{subequations}
\begin{align}
\label{eq8.3a}
\bullet\,& \text{the bulk } 0 < \xi<\pi\text{ of the twisted $\hat c=52$ open string}\\
\label{eq8.3b}
\bullet\,& \text{off mass-shell}\\
\label{eq8.3c}
\bullet\,& \text{as an open-string orbifold CFT}
\end{align}
\end{subequations}
where the {\it half-integer modeing} of standard orbifold theory persists 
for the open-string sector of the orientation orbifold.

Perhaps our most important conclusion however is the subtitle of this 
paper : ``Orientation orbifolds $\it include$ orientifolds''. This 
statement follows from the identification (8.2) and the work of Refs. 
 [17-19] where it is shown that, without appending any Chan-Paton 
structure, there are {\it many other orientation-orbifold string systems} 
$U(1)^{26}/(\mathbb{Z}_{2}(\mbox{w.s.})\times H)$ with higher 
fractional modeing $\{n(r)/\rho(\sigma)\}$. These constructions, with many twisted closed- and 
open-string sectors,  are  generically {\it inequivalent} [19] to untwisted string 
systems and should be further examined for consistency at the 
interacting level. 

Towards this, one should bear in mind 
that the twisted closed-string sectors of each orientation orbifold form
an essentially-ordinary space-time orbifold [1-15] at $c=26$, while each 
$\hat c=52$ twisted open-string sector has an equivalent but 
unconventionally twisted $c=26$ spectral 
description [19] in terms of the unconventional matter-field fractions $\{2n(r)/\rho(\sigma)\}$.
 For example, the orientation-orbifold CFT's 
\begin{subequations}
\begin{gather}
\label{eq8.4a}
\left(1,\,\omega_3,\,\omega^2_3\,;\,\, \tau_-,\, \tau_-\times \omega_3,
\,\tau_-\times\omega^2_3\right),\quad \omega_3^3=1 \\
\label{eq8.4b}
\left(1,\,\omega_4^2\,;\,\,\tau_-\times\omega_4,\,\tau_-\times\omega_4^3\right),
\quad\omega_4^4=1
\end{gather}
\end{subequations}
contain $1/3-$ and $1/4-$integer modeing respectively, though the 
$\hat c=52$ open strings of the latter have an equivalent 
half-integer-moded $c=26$ spectral description. It will be interesting 
in particular to understand whether the equivalent $c=26$ spectral 
description of the  general orientation orbifold-string can be extended 
(as seen for our special case here) to an equivalent but 
generically-new $c=26$ description of all these systems  at the interacting level. Some
 further remarks on the general-free 
bosonic orientation orbifold are included in the Appendix.

\newpage
\renewcommand{\thesection}{Appendix A.}
\renewcommand{\theequation}{A.\arabic{equation}}
\setcounter{equation}{0}
\section[\hspace*{12ex}Reducible Vertex Operators]{Reducible Vertex Operators\label{appendixA}}
For the particular orientation orbifold studied in this paper, Ref. [15] 
gives  a set of twisted vertex operators $\{\hat{g}(\mathcal{T})\}$ 
which have an additional $2\times2$ matrix structure relative to the 
ones studied  here. The matrix vertex operators are easily obtained from the vertex 
operators $\hat{\hat{g}}_{+}$ in Eq. (3.5) by the substitution
\begin{subequations}
\begin{gather}
\label{eqA.1a}
\hat x_{1\mu u}\to \hat x_{1\mu u}\tau_{u},\quad \bar u=0,1\\
\label{eqA.1b}
\hat{\hat{g}}_{+}(T) \,\to \,\,\,\,
\hat{g}(\mathcal{T}) \propto\,\,\,\, :e^{-\tfrac{i}{2}\hat x\cdot\mathcal{T}}:
\end{gather}
\end{subequations}
where $\tau_{0},\tau_{1}$ are respectively the $2\times 2$ 
unit matrix and the first Pauli matrix. In this notation, 
the twisted representation matrices $\{\mathcal{T}\}$ have the form
$\mathcal{T}_{\mu u}=T_{\mu} \tau_{u}$ where $\{T\}$ are the dimensionless 
closed-string momenta of the text.

Correspondingly, the commutators of the matrix emission operators $\hat{g}(\mathcal{T})$
at $\xi=\pi$ with the twisted currents $\hat{J}$ and extended Virasoro generators 
$\hat{L}$ are obtained from those in Eq. (3.9) by the additional substitutions
\begin{equation}
\label{eqA.2}
\hat g_+(T)\to \hat g (\mathcal{T}) ,\quad\, 
\hat J_{1\mu u}\to \tau_u\hat J_{1\mu u},\quad\,
\hat L_u\to \tau_u\hat L_u.
\end{equation}
Twisted matrix tree graphs $\hat{A}(\{\mathcal{T}\})$ can also be 
constructed as in Eq. (4.1), now using matrix multiplication of 
$\hat{g}$'s with all $\tau^{(i)}_u=\tau_u$. Then one finds that the 
following matrix gauges
\begin{equation}
\label{eqA.3}
\hat W_u(m+\tfrac{u}{2})= e^{-i\pi(m+\tfrac{u}{2})}
\tau_u \hat L_u(m+\tfrac{u}{2})-
\left(\hat L_0(0)+m+\tfrac{u}{2}-\hat a_2\right)
\end{equation}
are operative in the matrix trees, leading to the same extended 
physical state conditions (5.4b). Explicit evaluation of the matrix 
trees gives
\begin{equation}
\label{eqA.4}
\hat A_n(\{{\cal T}\})=\tau_0 \hat A_n(\{T\})
\end{equation}
that is, two copies of the NN amplitudes $\hat A_n(\{T\})$ obtained 
in Eq. (6.4).

This result can be understood as reducibility of the matrix vertex 
operators $\hat{g}(\mathcal{T})$
\begin{subequations}
\begin{gather}
\label{eqA.5a}
U=\frac{1}{\sqrt{2}}
\begin{pmatrix}
1 & 1\\
1 & -1 
\end{pmatrix},\quad\,
U \tau_1 U^\dagger=\tau_3\\
\label{eqA.5b}
U \hat g({\cal T}) U^\dagger=
\begin{pmatrix}
\hat g_+(T) & 0\\
0 & \hat g_-(T)
\end{pmatrix}
\end{gather}
\end{subequations}
where $\hat{g}_{+}(T)$ (or $\hat{\hat{g}}_{+}$ for all $\xi)$ is the vertex operator of the text.
The vertex operator  $\hat{g}_{-}(T)$ differs from $\hat{g}_{+}(T)$ 
only by a sign reversal $\hat x_{1\mu 1}\to -\hat x_{1\mu 1}$ of the 
NN component of the $\hat{c}=52$ string, which reproduces Eq. (A.4) because all vertex-operator 
contractions are pairwise.

In the computations of the text, I have kept only one irreducible component $\hat{g}_{+}$, 
which is equivalent to the replacement $\tau_{u}\to 1, \mathcal{T}\to T$ in the matrix 
vertex operator $\hat{g}(\mathcal{T})$. Certainly, this choice is sufficient to 
satisfy ordinary $\mbox{open}\leftrightarrow\mbox{closed}$ string 
duality in this case. More physically, the prescription may be 
understood as division by the symmetry  $\hat g_+\leftrightarrow 
\hat g_-$, which is a residual form of world-sheet parity [12,13,15] in this basis. 

Refs. [12,15] give mode expansions for the coordinates 
$\hat{x}_{\sigma}$ and vertex-operator equations
for the twisted vertex operators
\begin{equation}
\label{eqA.6}
\hat{g}_(\mathcal{T},\sigma)\propto\,\,\,\,
:e^{i\hat{x}_{\sigma}^{n(r)\mu u}\mathcal{T}_{n(r)\mu}(T,\sigma)\bigotimes\tau_{u}}:
\end{equation}
in open-string sector $\sigma$ of the general free-bosonic orientation 
orbifold (1.1). The explicit form of the twisted representation matrices 
$\{\mathcal{T}(T,\sigma)\}$ is given in Eq. (2.20d) of Ref. [12], 
where the quantities $\{T\}$ in this application are the same dimensionless critical 
closed-string momenta of the text. Each of these vertex operators has the
 same $2\times2$ reducible matrix structure 
discussed here and, dividing by world-sheet parity, I would speculate that the same prescription 
$\tau_{u}\to 1$ will be sufficient to satisfy $\mbox{open}\leftrightarrow\mbox{closed}$
string duality in the general case.

\section*{Acknowledgments}

For helpful information,  discussions and encouragement, I thank L. 
Alvarez-Gaum$\acute{e}$, K. Bardakci, I. Brunner,
J. de Boer, D. Fairlie, O. Ganor, E. Gimon, C. Helfgott, E. Kiritsis, R. 
Littlejohn, S. Mandelstam, J. McGreevy, N. Obers, A. Petkou, E. 
Rabinovici, V. Schomerus, K. Schoutens, C. Schweigert and E. Witten. This work was 
supported in part by the Director, Office of Energy Research, Office of 
High Energy and Nuclear Physics, Division of High Energy Physics of the U.S.
Department of Energy under Contract DE-AC02-O5CH11231 and in part by the National
Science Foundation under grant PHY00-98840.

\end{document}